\documentclass[12pt]{article}
\usepackage{amsmath,amsfonts}
\usepackage[nosort]{cite}
\usepackage{dsfont}

\def\be{\begin{equation}}
\def\ee{\end{equation}}
\def\bea{\begin{eqnarray}}
\def\eea{\end{eqnarray}}

\newcommand{\vev}[1]{{\left< {#1} \right>}}

\newcommand{\Tr}{{\rm Tr\,}}

\newcommand{\cA}{{\mathcal A}}

\newcommand{\cN}{{\mathcal N}}
\newcommand{\cP}{{\mathcal P}}

\newcommand{\cS}{{\mathcal S}}

\renewcommand{\title}[1]{\vbox{\center\LARGE{#1}}\vspace{5mm}}
\renewcommand{\author}[1]{\vbox{\center#1}\vspace{5mm}}
\newcommand{\address}[1]{\vbox{\center\em#1}}
\newcommand{\email}[1]{\vbox{\center\tt#1}\vspace{5mm}}

\begin{document}
\begin{titlepage}
\begin{center}
\begin{flushright}
HU-EP-07/23\\
YITP-SB-07-23\\
Imperial/TP/07/RR/03\\
{\tt arXiv:0707.2699}
\end{flushright}
\vskip .5cm

\title{Wilson loops: From four-dimensional SYM\\
to two-dimensional YM}

\author{Nadav Drukker$^{1,a}$,
Simone Giombi$^{2,b}$,
Riccardo Ricci$^{3,4,c}$,
Diego Trancanelli$^{2,d}$}

\address{$^1$Humboldt-Universit\"at zu Berlin, Institut f\"ur Physik,\\
Newtonstra{\ss}e 15, D-12489 Berlin, Germany\\
\medskip
$^2$C. N. Yang Institute for Theoretical Physics,\\
State University of New York at Stony Brook\\
Stony Brook, NY 11794-3840, USA\\
\medskip
$^3$ Theoretical Physics Group, Blackett Laboratory,\\
Imperial College, London, SW7 2AZ, U.K.\\
\medskip
$^4$ The Institute for Mathematical Sciences, \\
Imperial College, London, SW7 2PG, U.K.}

\email{$^a$drukker@physik.hu-berlin.de,
$^b$sgiombi@max2.physics.sunysb.edu,
$^c$r.ricci@imperial.ac.uk,
$^d$dtrancan@max2.physics.sunysb.edu}

\end{center}

\abstract{
\noindent
In this note we study supersymmetric Wilson loops restricted to an $S^2$
submanifold of four-dimensional space in $\cN=4$ super Yang-Mills.
We provide evidence from both perturbation theory and the $AdS$ dual
that those loops are equal to the analogous observables in two-dimensional
Yang-Mills on $S^2$ (excluding non-perturbative contributions).
This relates a subsector of $\cN=4$ SYM to a low-dimensional soluble
model and also suggests that this subsector of $\cN =4$ SYM is invariant under
area preserving diffeomorphisms.
}

\vfill

\end{titlepage}


Maximally supersymmetric Yang-Mills (SYM) theory in four
dimensions is a remarkable theory which is conformally invariant,
is conjectured to be dual to string theory in $AdS_5\times S^5$
and seems to be integrable in the planar limit. This last property
was uncovered by relating the anomalous dimensions of local
operators with the Hamiltonian of a 1-dimensional spin-chain,
which turned out to be integrable \cite{Minahan:2002ve}. In this
note we suggest another connection between operators in $\cN=4$
SYM and a low-dimensional integrable model: Certain supersymmetric
Wilson loops in this theory and Wilson loops in two-dimensional
purely bosonic YM.

Since $\cN=4$ SYM is a very rich theory, a better understanding of
it may be gained by restricting to some simpler subsector. In the
example of the local operators the calculation evolved through the
different closed subsectors, like the one comprising of only two
scalars which is identical at one-loop to the Heisenberg XXX
spin-chain, to the full supersymmetric chain. Still most
calculations are valid only for infinitely long operators, but
there has been continuous progress towards solving more and more
complicated problems.

In the study of Wilson loop operators in $\cN=4$ SYM much less advance
has been made. Some family of supersymmetric operators
\cite{Zarembo:2002an} seems to be completely trivial, with evidence
from both the string theory side and from perturbation theory
\cite{Guralnik:2003di,Guralnik:2004yc,Dymarsky:2006ve}.
Otherwise there are proposed all-loop results only for the $1/2$
BPS circle \cite{Erickson:2000af,Drukker:2000rr}
and some recently found $1/4$ BPS examples \cite{Drukker:2006ga}.
In this note we propose an all-loop result for a family of $1/8$ BPS
Wilson loops following an arbitrary curve on an $S^2$ in space-time.
Our hope is that this connection, beyond allowing the calculation of
these specific operators, will also shed some light on hidden symmetries
of the four dimensional gauge theory.

The operators we consider are a subclass of those constructed recently in
\cite{Drukker:2007dw} (more details will be provided in \cite{next}).
For an arbitrary curve $x^\mu(s)$ on a 2-sphere of radius $R$ we define
\begin{equation}
W=\frac{1}{N}\Tr\,\cP\exp \int
\left(iA_\mu+\frac{x^\nu\epsilon_{\nu\rho\mu}}{R}
M^\rho{}_I\Phi^I \right)dx^\mu\,.
\label{susy-loop}
\end{equation}
Here $\epsilon_{\nu\rho\mu}$ is the three index anti-symmetric tensor,
the matrix $M^\rho{}_I$ satisfies $M^\rho_IM^\nu_I=\delta^{\rho\nu}$
and $M^\rho_IM^\rho_J=\delta_{IJ}$. It serves to choose
three of the six real scalar fields, say $\Phi^1$, $\Phi^2$ and $\Phi^3$
to couple to the loop.

In \cite{Drukker:2007dw} we showed that those operators, for a generic
curve, preserve four supercharges which are a linear combination of the
super-Poincar\'e generators (the ``$Q$''s) and super-conformal
generators (the ``$S$''s). Here we wish to evaluate their expectation values.

Consider such a loop in perturbation theory expanding to second
order, the contraction of the gauge fields and scalars gives
\begin{equation}
\vev{W}=1-\frac{1}{2N}\Tr\int dx^\mu\,dy^\nu
\left(\vev{A_\mu(x)\,A_\nu(y)}-
\frac{x^\sigma y^\eta\epsilon_{\sigma\rho\mu}
\epsilon_{\eta\zeta\nu}}{R^2}M^\rho{}_IM^\zeta{}_J
\vev{\Phi^I(x)\,\Phi^J(y)}\right)\,.
\end{equation}
Using the property of $M$ and that
$\epsilon_{\sigma\rho\mu}\epsilon_{\eta\rho\nu}
=\delta_{\sigma\eta}\delta_{\mu\nu}-
\delta_{\sigma\nu}\delta_{\mu\eta}$
we find in the Feynman gauge
\begin{equation}
\vev{W}=1-\frac{1}{2N}\Tr(T^aT^b)\int dx^\mu\,dy^\nu\,
\frac{g_{4d}^2\,\delta^{ab}}{4\pi^2R^2}
\left(\frac{1}{2}\delta_{\mu\nu} -\frac{(x-y)_\mu
(x-y)_\nu}{(x-y)^2}\right)\,. \label{eff-prop}
\end{equation}
$T^a$ are the generators of the gauge group, say $U(N)$, and satisfy
$\Tr(T^aT^a)=N^2/2$.

Instead of proceeding to evaluate the integral, we would like to examine it
in this form. Recall that for the circular $1/2$ BPS Wilson loop the
combination of the scalar and gauge field propagators was a constant,
which led to the identification of that operator with the zero-dimensional
matrix model \cite{Erickson:2000af,Drukker:2000rr}.
In our case, with an arbitrary curve on $S^2$, the
propagator is not a constant, instead it involves an effective coupling
$g_{4d}^2/4\pi^2R^2$ of dimension 2 and a dimensionless
configuration-space propagator. Those are appropriate for a gauge field
in two dimensions and indeed they can be identified with the coupling
and propagator of pure Yang-Mills in two dimensions on $S^2$.

To see that this is a vector propagator on $S^2$
we will change coordinates and parameterize the sphere in terms
of complex coordinates $z$ and $\bar z$ as
\begin{equation}
\vec x = \frac{R}{1+z\bar z}\left(z+\bar z,\, -i(z-\bar
z),\,1-z\bar z\right)\,. \label{complex-coords}
\end{equation}
In these coordinates, the $S^2$ metric takes the standard
Fubini-Study form
\begin{equation}
ds^2 = \frac{4R^2 \,dz d\bar z}{(1+ z \bar z)^2}\,. \label{metric}
\end{equation}

For the gauge theory on the sphere consider the generalized Feynman
gauge with gauge
parameter $\xi=-1$, so the kinetic action is
\begin{equation}
L=\frac{\sqrt{g}}{g_{2d}^2} \left[\frac{1}{4}(F_{ij}^a)^2
-\frac{1}{2}(\nabla^iA^a_i)^2\right] = -\frac{\sqrt{g}}{g_{2d}^2}
(g^{z \bar z})^2\left[ (\nabla_zA_{\bar z}^a)^2+(\nabla_{\bar
z}A_z^a)^2\right]\, ,
\end{equation}
where in the last equality we have ignored interaction terms, and
the covariant derivatives are taken with respect to the metric
(\ref{metric}). A simple calculation shows that the propagators
\begin{equation}
\begin{aligned}
\Delta^{ab}_{zz}(z,w)&
=R^2\delta^{ab}\frac{g_{2d}^2}{\pi}\frac{1}{(1+ z \bar z)}\frac{1}{(1+w \bar w)}\frac{\bar z-\bar w}{z-w}\, ,\\
\Delta^{ab}_{\bar z\bar z}(z,w)& =R^2\delta^{ab}\frac{g_{2d}^2}{\pi}
\frac{1}{(1+ z \bar z)}\frac{1}{(1+w \bar w)} \frac{z-w}{\bar
z-\bar w}\, ,
\label{props}
\end{aligned}
\end{equation}
satisfy
\begin{equation}
\frac{2}{g_{2d}^2}(g^{z \bar z})^2 \nabla^2_{\bar z}
\Delta^{ab}_{zz}(z,w) =\delta^{ab}
\frac{1}{\sqrt{g}}\delta^2(z-w)\, , \label{norma}
\end{equation}
and similarly for $\Delta_{\bar z\bar z}$.
By doing the change of variables to the complex coordinates
(\ref{complex-coords}), one can then see that the effective
propagator in (\ref{eff-prop}) agrees with the 2d vector
propagators (\ref{props}) when the 2d and 4d couplings are related
by
\begin{equation}
g_{2d}^2 = -\frac{g_{4d}^2}{4\pi R^2}\,.
\end{equation}

Another way to show that (\ref{eff-prop}) may serve as a propagator on
$S^2$ is to consider a gauge invariant source, i.e. a Wilson loop in the
Abelian theory. The field strength that is derived from it should solve
the equations of motion
\begin{equation}
\frac{1}{\sqrt{g}}\partial_i\left(\sqrt{g}F^{ij}\right)=0\,.
\end{equation}
Therefore the field-strength should be a constant unless one crosses the
Wilson loop.

Given a source along the curve $y$, using the effective propagator on $S^2$,
the gauge field at $x$ is
\begin{equation}
A_\mu=\frac{g_{2d}^2}{\pi}\int dy^\nu
\left(\frac{1}{2}\delta_{\mu\nu}
-\frac{(x-y)_\mu (x-y)_\nu}{(x-y)^2}\right)\,,
\end{equation}
and the resulting field-strength, gotten by differentiation and
projection in the directions tangent to the sphere, is
\begin{equation}
F_{\mu\nu}=-\frac{g_{2d}^2}{\pi}\int ds\,
\frac{-\dot y_\mu y_\nu+\dot y_\nu y_\mu}{(x-y)^2}\,.
\end{equation}
The associated dual scalar reads
\begin{equation}
\tilde F=-\frac{g_{2d}^2}{\pi R}\int ds\,
\frac{\epsilon_{\mu\nu\rho}\,\dot y^\mu y^\nu x^\rho}{(x-y)^2}\,.
\end{equation}
To evaluate $\tilde{F}$ explicitly we define $\theta(s)$ to be the
angle between the points $x$ and $y$. Then the numerator is
proportional to the one-form normal to $d\theta$, which we label
by $d\phi$. This gives
\begin{equation}
\tilde F=\frac{g_{2d}^2}{\pi}\int d\phi\,
\frac{\sin^2\theta}{2(1-\cos\theta)} =\frac{g_{2d}^2}{\pi}\int
d\phi\,\cos^2\frac{\theta}{2}
=\frac{g_{2d}^2}{2\pi R^2}\int_{\Sigma_2} d\theta\,d\phi\sin\theta
=2 g_{2d}^2\frac{\cA_2}{\cA}\,,
\end{equation}
where $\cA_2$ is the area of the part of the sphere enclosed by
the loop and not including $x$ and $\cA$ the total area. Clearly
this is a constant unless $x$ crosses the loop. Using this it is
simple to evaluate the Wilson loop at the quadratic order using
Stokes theorem for the $x$ integral in (\ref{eff-prop}). We get
\begin{equation}
\vev{W}=1-\frac{N}{4}\int_{\Sigma_1}\tilde F
=1-g_{2d}^2N\,\frac{\cA_1\cA_2}{2\cA}\,,
\label{g2-result}
\end{equation}
and the result is the product of the areas of the two parts of the sphere
separated by the loop and it clearly does not depend on the order
of the $y$ and $x$ integrals.

We were not able to calculate higher order graphs for loops of
arbitrary shape, neither in four dimensions, nor in two. Note that
as opposed to the light-cone gauge, in this gauge there are
interactions in two dimensions, so the calculation is non-trivial.
But two-dimensional YM is a soluble theory \cite{Migdal:1975zg,
Rusakov:1990rs}, so we can use known results (derived by other
methods) and compare them to some results in four dimensions,
including some strong coupling results from the $AdS$ dual of
$\cN=4$.

The above perturbative calculation of the Wilson loop in two dimensions is
very similar to the one performed by Staudacher and Krauth in
\cite{Staudacher:1997kn}.
There the calculation is in the plane and using the light-cone gauge,
while we have the calculation on $S^2$ and employ a different gauge.
Still we checked for the circular loop that both prescriptions in two
dimensions lead to the same result, and as we will show below, in
that case it agrees also with the four-dimensional expressions.

The important part about the calculation in
\cite{Staudacher:1997kn} is not the choice of gauge, but the
choice of regularization prescription of a pole in the derivation
of the configuration-space propagator. The one they used was
proposed by Wu, Mandelstam and Leibbrandt
(WML) \cite{Wu:1977hi,Mandelstam:1982cb,Leibbrandt:1983pj} and it
can be used also in Euclidean signature space. One may change to
the light-cone gauge on the sphere by taking $A_{\bar z}=0$ and
then, using the same prescription the propagator for $A_z$ would
be double the one in (\ref{props}). Their propagator is the flat
space limit of that, and using it they are able to sum up all the
ladders and find that the Wilson loop is given by
\begin{equation}
\vev{W}=\frac{1}{N}L_{N-1}^1\left(g_{2d}^2\cA_1\right)
\exp\left[-\frac{g_{2d}^2\cA_1}{2}\right]\,,
\label{2d-result}
\end{equation}
where $L_{N-1}^1$ is a Laguerre polynomial and $\cA_1$ is the area enclosed
by the loop. This is equal to the expectation value of a Wilson loop in the
Gaussian Hermitean matrix model.

This expression has an obvious generalization to $S^2$ with the simple
replacement $\cA_1\to \cA_1\cA_2/\cA$, where the combination of the areas is the same as appeared
in (\ref{g2-result}).

The reader may be puzzled by those formulas, since they do not agree
with the exact solution of YM in two dimensions. This confusion was
resolved by Bassetto and Griguolo \cite{Bassetto:1998sr}, who showed
that (\ref{2d-result}) may be extracted from
the exact result by restricting to the zero instanton sector following the
expansion of \cite{Witten:1992xu}. It was therefore concluded that
the perturbative calculation of \cite{Staudacher:1997kn}, using the
light-cone gauge and the WML prescription
for performing the momentum integrals does not capture non-perturbative
effects.

The two dimensional propagator we found is not in the same gauge,
but it also is defined by the WML prescription. Since we expect
the result not to depend on gauge, we conclude that the result of
the perturbative 2-dimensional YM sum that our four-dimensional
Wilson loops seem to point to is given by
\begin{equation}
\vev{W}
=\frac{1}{N}L_{N-1}^1\left(-g_{4d}^2\,\frac{\cA_1\cA_2}{\cA^2}\right)
\exp\left[\frac{g_{4d}^2}{2}\,\frac{\cA_1\cA_2}{\cA^2}\right]\,.
\label{4d-result}
\end{equation}
The expansion of this
expression to order $g_{4d}^2$ agrees with the aforementioned result
(\ref{g2-result}). In the remainder of this note we will provide further
evidence that this expression correctly captures the
Wilson loops in four dimensions.

The most comprehensive test of this conjecture comes from considering
latitude lines on $S^2$. In \cite{Drukker:2006ga,Drukker:2006zk}
evidence was given that the curve at
latitude $\theta$ is equal to (\ref{4d-result}) with the replacement
$\cA_1\cA_2/\cA^2\to\frac{1}{4}\sin^2\theta$. For a latitude
the areas bound by the curve are
\begin{equation}
\cA_1=2\pi R^2(1-\cos\theta)\,,\qquad
\cA_2=2\pi R^2(1+\cos\theta)\,,
\end{equation}
so indeed $\cA_1\cA_2=\frac{1}{4}\cA^2\sin^2\theta$ and the above
expression agrees
with the results found there which were supported by perturbative calculations,
classical strings in $AdS_5\times S^5$ as well as a D3-brane calculation.

As a special case, when $\theta=\pi/2$, this includes the $1/2$ BPS
circle of \cite{Erickson:2000af,Drukker:2000rr}.

There is another family of examples where we could test this expression in
the large $N$ and large $g_{4d}^2N$ limit, where (\ref{4d-result}) reduces
to
\begin{equation}
\vev{W}
\sim\frac{\cA}{\sqrt{g_{4d}^2N\cA_1\cA_2}}
I_1\left(\frac{2\sqrt{g_{4d}^2N \cA_1\cA_2}}{\cA}\right)
\sim\exp\left(\frac{2\sqrt{g_{4d}^2N \cA_1\cA_2}}{\cA}\right)\,,
\label{bessel}
\end{equation}
with $I_1$ a modified Bessel function.

The examples we studied are for two longitude lines connected at the north
and south pole of $S^2$ and separated by an angle $\theta$.
In perturbation theory we calculated it only to order $g_{4d}^2$ as we did
above for the general curve (\ref{g2-result}). But in this case we can
also do the calculation in $AdS$ \cite{next}, where the curve may be
mapped by a stereographic projection to a cusp in the plane and then
calculated by generalizing \cite{Drukker:1999zq}. The result of
the calculation is
\begin{equation}
\cS=-\frac{\sqrt{\theta(2\pi-\theta)g_{4d}^2N}}{\pi}\,.
\end{equation}
The expectation value of the Wilson loop is the exponent of minus the
classical action which, noting that the areas are
\begin{equation}
\cA_1=2\theta\,,\qquad
\cA_2=2(2\pi-\theta)\,.
\end{equation}
agrees again with large coupling limit of the two-dimensional
YM calculation.

We have tried in this note to provide evidence that the supersymmetric
Wilson loops of \cite{Drukker:2007dw} on an $S^2$ are equal to
the usual Wilson loops in 2-dimensional Yang-Mills in the
Wu-Mandelstam-Leibbrandt prescription. The leading terms in the
perturbative calculation agree as do some all-order expressions as
well as other examples where the strong coupling result is known from
an $AdS$ calculation. We note however that we have not been able to
substantiate this correspondence beyond the leading order calculation
and those examples, so it is conceivable that the two dimensional theory
describing those loops is more complicated, with the same kinetic term
as YM, but with more complicated (potentially also non-local)
interactions.

If this correspondence holds, it would be one of those miracles
of $\cN=4$ SYM, where there seems
to be a ``consistent truncation'' to the sphere and we
can simply ignore all the fields away from it. It is also quite remarkable
since YM in 2d is invariant under area preserving diffeomorphisms. So a
subsector of the superconformal theory is invariant under all transformations
which change angles but keep areas constant. It would be interesting to
find out if those properties manifest themselves in a deeper way in the entire
theory beyond this subsector.


\subsection*{Acknowledgments}
N.D. would like foremost to thank Matthias Staudacher and  Poul Olesen for
pointing out the similarities between the circular Wilson loop in 4 and 2
dimensions and who partially inspired this work.
In addition we are happy to thank Sunny Itzhaki and Jan Plefka
for interesting discussions and Luca Griguolo for correspondence.
N.D. is grateful to the University of Barcelona and the Galileo Galilei
Institute for their hospitality in the course of this work and to the
INFN for partial financial support.
S.G. and D.T.  acknowledge partial financial support through the
NSF award PHY-0354776.



\begin{thebibliography}{20}

\bibitem{Minahan:2002ve}
  J.~A.~Minahan and K.~Zarembo,
  ``The Bethe-ansatz for $\cN = 4$ super Yang-Mills,''
  JHEP {\bf 0303} (2003) 013
  [hep-th/0212208].

\bibitem{Zarembo:2002an}
  K.~Zarembo,
  ``Supersymmetric Wilson loops,''
  Nucl.\ Phys.\  B {\bf 643}, 157 (2002)
  [hep-th/0205160].

\bibitem{Guralnik:2003di}
Z.~Guralnik and B.~Kulik,
``Properties of chiral Wilson loops,''
JHEP {\bf 0401}, 065 (2004)
[hep-th/0309118].

\bibitem{Guralnik:2004yc}
Z.~Guralnik, S.~Kovacs and B.~Kulik,
``Less is more: Non-renormalization theorems from lower dimensional
superspace,''
Int.\ J.\ Mod.\ Phys.\  A {\bf 20}, 4546 (2005)
[hep-th/0409091].

\bibitem{Dymarsky:2006ve}
A.~Dymarsky, S.~Gubser, Z.~Guralnik and J.~M.~Maldacena,
``Calibrated surfaces and supersymmetric Wilson loops,''
hep-th/0604058.

\bibitem{Erickson:2000af}
J.~K.~Erickson, G.~W.~Semenoff and K.~Zarembo,
``Wilson loops in $\cN = 4$ supersymmetric Yang-Mills theory,''
Nucl.\ Phys.\ B {\bf 582}, 155 (2000)
[hep-th/0003055].

\bibitem{Drukker:2000rr}
N.~Drukker and D.~J.~Gross,
``An exact prediction of $\cN = 4$ SUSYM theory for string theory,''
J.\ Math.\ Phys.\  {\bf 42}, 2896 (2001)
[hep-th/0010274].

\bibitem{Drukker:2006ga}
N.~Drukker,
``$1/4$ BPS circular loops, unstable world-sheet instantons and the matrix
model,''
JHEP {\bf 0609}, 004 (2006)
[hep-th/0605151].

\bibitem{Drukker:2007dw}
N.~Drukker, S.~Giombi, R.~Ricci and D.~Trancanelli,
``More supersymmetric Wilson loops,''
arXiv:0704.2237.

\bibitem{next}
N.~Drukker, S.~Giombi, R.~Ricci and D.~Trancanelli, in preparation.

\bibitem{Migdal:1975zg}
  A.~A.~Migdal,
  ``Recursion equations in gauge field theories,''
  Sov.\ Phys.\ JETP {\bf 42} (1975) 413
  [Zh.\ Eksp.\ Teor.\ Fiz.\  {\bf 69} (1975) 810].

\bibitem{Rusakov:1990rs}
  B.~E.~Rusakov,
  ``Loop averages and partition functions in $U(N)$ gauge theory on
  two-dimensional manifolds,''
  Mod.\ Phys.\ Lett.\  A {\bf 5} (1990) 693.

\bibitem{Staudacher:1997kn}
M.~Staudacher and W.~Krauth,
``Two-dimensional {QCD} in the Wu-Mandelstam-Leibbrandt prescription,''
Phys.\ Rev.\  D {\bf 57}, 2456 (1998)
[hep-th/9709101].

\bibitem{Wu:1977hi}
  T.~T.~Wu,
  ``Two-dimensional Yang-Mills theory in the leading $1/N$ expansion,''
  Phys.\ Lett.\  B {\bf 71} (1977) 142.

\bibitem{Mandelstam:1982cb}
  S.~Mandelstam,
  ``Light cone superspace and the ultraviolet finiteness of the $\cN=4$
  model,''
  Nucl.\ Phys.\  B {\bf 213} (1983) 149.

\bibitem{Leibbrandt:1983pj}
  G.~Leibbrandt,
  ``The light cone gauge in Yang-Mills theory,''
  Phys.\ Rev.\  D {\bf 29} (1984) 1699.

\bibitem{Bassetto:1998sr}
  A.~Bassetto and L.~Griguolo,
  ``Two-dimensional {QCD}, instanton contributions and the perturbative
  Wu-Mandelstam-Leibbrandt prescription,''
  Phys.\ Lett.\  B {\bf 443}, 325 (1998)
  [hep-th/9806037].

\bibitem{Witten:1992xu}
E.~Witten,
``Two-dimensional gauge theories revisited,''
J.\ Geom.\ Phys.\  {\bf 9}, 303 (1992)
[hep-th/9204083].

\bibitem{Drukker:2006zk}
N.~Drukker, S.~Giombi, R.~Ricci and D.~Trancanelli,
``On the D3-brane description of some 1/4 BPS Wilson loops,''
JHEP {\bf 0704}, 008 (2007)
[hep-th/0612168].

\bibitem{Drukker:1999zq}
N.~Drukker, D.~J.~Gross and H.~Ooguri,
``Wilson loops and minimal surfaces,''
Phys.\ Rev.\  {\bf D60}, 125006 (1999)
[hep-th/9904191].

\end{thebibliography}
\end{document}